\documentclass[
 aps, pra,
 amsmath,amssymb,
 11pt,
 final,
tightenlines,
 twoside,
 twocolumn,
 nofloats,
nofootinbib,
 superscriptaddress,
showkeys,
showkeywords,
 ]
{revtex4-2}

\usepackage[T2A]{fontenc}
\usepackage[utf8x]{inputenc}
\usepackage[russian,english]{babel}
\usepackage{graphicx}
\usepackage{dcolumn}
\usepackage{bm}

\input{maik.rty}

\setcitestyle{authoryear,round}
\def\squareforqed{\hbox{\rlap{$\sqcap$}$\sqcup$}}

\def\sq{\ifmmode\squareforqed\else{\unskip\nobreak\hfil
\penalty50\hskip1em\null\nobreak\hfil\squareforqed
\parfillskip=0pt\finalhyphendemerits=0\endgraf}\fi}

\def\utw{\smash{\rlap{\lower5pt\hbox{$\sim$}}}}

\def\udtw{\smash{\rlap{\lower6pt\hbox{$\approx$}}}}

\def\diameter{{\ifmmode\mathchoice
{\ooalign{\hfil\hbox{$\displaystyle/$}\hfil\crcr
{\hbox{$\displaystyle\mathchar"20D$}}}}
{\ooalign{\hfil\hbox{$\textstyle/$}\hfil\crcr
{\hbox{$\textstyle\mathchar"20D$}}}}
{\ooalign{\hfil\hbox{$\scriptstyle/$}\hfil\crcr
{\hbox{$\scriptstyle\mathchar"20D$}}}}
{\ooalign{\hfil\hbox{$\scriptscriptstyle/$}\hfil\crcr
{\hbox{$\scriptscriptstyle\mathchar"20D$}}}}
\else{\ooalign{\hfil/\hfil\crcr\mathhexbox20D}}%
\fi}}

\begin{document}

\selectlanguage{english}



\title{Observations of Galaxies at $z\gtrsim10$ Allow to Test Cosmological Models with Features in the Initial Power Spectrum}

\author{\firstname{S.~V.}~\surname{Pilipenko}}
\email{spilipenko@asc.rssi.ru}
\affiliation{P.N. Lebedev Physical Institute, Astro Space Center, Moscow, 117997 Russia}

\author{\firstname{S.~A.}\surname{Drozdov}}
\affiliation{P.N. Lebedev Physical Institute, Astro Space Center, Moscow, 117997 Russia}

\author{\firstname{M.~V.}\surname{Tkachev}}
\affiliation{P.N. Lebedev Physical Institute, Astro Space Center, Moscow, 117997 Russia}

\author{\firstname{A.~G.}~\surname{Doroshkevich}}
\affiliation{P.N. Lebedev Physical Institute, Astro Space Center, Moscow, 117997 Russia}
\affiliation{National research centre Kurchatov institute, Moscow, Russia}

\begin{abstract}
The initial power spectrum of density perturbations, generated during the inflationary epoch, is now constrained by observations on scales $\lambda>5$~Mpc and has a power-law form. The peculiarities of the inflationary process can lead to the appearance of non-power-law contributions to this spectrum, such as peaks. The exact size and shape of the peak cannot be predicted in advance. In this paper, we propose methods for searching for such peaks in the region of the spectrum with $\lambda<5$~Mpc. Perturbations on these scales enter the nonlinear stage at $z\gtrsim10$, which is now becoming accessible to observations. Our studies of numerical models of large-scale structure with peaks in the initial spectrum have shown that spectral features on scales with $\lambda>0.1$~Mpc manifest in the clustering of galaxies, as well as affect their mass function, sizes, and density. Studying these characteristics of distant galaxies will allow us to constrain cosmological models with peaks.
\end{abstract}

\maketitle

\section{INTRODUCTION}
In the currently accepted cosmological model, the initial power spectrum of density perturbations has a {power-law form}, $P\propto k^{0.961}$ \citep{Planck18_VI}. According to observations of the relic radiation and the distribution of galaxies in space, this spectrum is verified up to a scale of $k\approx1$~$h$/Mpc\footnote{Here and below $h=H/100$~km/s/Mpc, where $H$ is the Hubble constant} \citep{Chabanier19}. Although such a spectrum is obtained even in the simplest inflation models, there are more complex models in which the spectrum may not have a power-law shape. {For example, as shown in the work \cite{Ivanov94}, a peak arises in the presence of a <<step>> in the inflaton potential.} An overview of models {of inflation leading to the appearance of peaks of various shapes} is given in the article \cite{2023JCAP...04..011I}, {where 42 references are given to works that considered 8 scenarios of inflation leading to the formation of peaks. Let us briefly consider some of them.}

{Models of hybrid inflation are developed to solve the problem of too large a starting value of the field for the most popular model of chaotic inflation. In these models, there may be overproduction of primordial black holes (PBHs) above the observed limits, but by adjusting the parameters it is possible to remove the excess of PBHs, but peaks in the power spectrum may remain \citep{GraciaBellido96}. The model of <<new inflation>> \citep{Linde82,Albrecht82} is attractive from the point of view of supergravity theory, but in this theory there is also the problem of too large a starting value of the field. It is proposed to solve it by means of <<double inflation>>, i.e., by adding a pre-inflationary stage before the main inflationary expansion. A change in the rate of expansion of the Universe during the transition from one inflation to another in such models leads either to the birth of PBHs or to the appearance of a peak in the power spectrum \citep{Kawasaki98}. The presence in the early Universe of fields other than the inflaton will lead to the generation of entropic perturbations, which can also cause the appearance of peaks. One of the possible candidates is the axion field, interesting because axions can make up dark matter in our Universe \citep{Ando18}.} The power spectrum can also acquire a peak in Brans-Dicke gravity theory \citep{Sletmoen24} or in models with a primordial magnetic field, see, for example, \cite{Ralegankar24}. {The primordial magnetic field induces density perturbations in the baryonic medium with a spectrum that increases towards small scales \citep{Gopal03,Pandey13}, but only for scales exceeding the magnetic Jeans scale \citep{Kim96,Kahniashvili10}. This leads to the formation of a <<hump>> on the complete power spectrum of matter.}

{Until recently, cosmological models with peaks in the spectrum did not have much popularity because they require the introduction of new parameters into cosmology. However, the discovery of a large number of bright galaxies at redshifts $z>10$ on the James Webb Space Telescope (JWST) \citep{Naidu2022b,Castellano22,Finkelstein22,Donnan23,Labbe23} has caused a number of scientists to question the validity of the standard cosmological model \citep{Boylan-Kolchin23,Lovell22}. The question of whether JWST data can be explained without changes in the cosmological model is debatable \citep{Chen23,Prada23,Shen23}, and makes us think about possible complications of the model. One such complication is the increase in the amplitude of perturbations of the power spectrum at small scales, corresponding to wave numbers $k>1$ $h$/Mpc, which was considered in the works \citep{Parashari23,Padmanabhan_2023,Tkachev23}.}

{Increasing the power of Gaussian density perturbations in a certain range of scales leads to the earlier formation of structures with masses corresponding to this range of scales. This is easy to see from the Press-Schechter model \citep{press}, in which spherical dark matter halos are formed when perturbations reach a certain critical value. The earlier the halos are formed, the more compact and denser they are, due to the fact that during virilization the density of the halo increases by several hundred times compared to the average density of the Universe at the time of halo formation. Press-Schechter theory and its refinements \citep{Sheth99,Despali_2016} allow us to calculate the halo mass function for a given spectrum. For spectra with a peak, this was done in \cite{Padmanabhan_2023}, as well as using numerical simulation in \cite{Knebe01} and \cite{Tkachev23}. More dense and earlier formed halos in the model with a peak can make up the population of objects that are observed at high $z$ on the JWST telescope.}

{To compare the results of observations with cosmological models, it is necessary to take into account that} now for distant galaxies they determine not the dynamic mass, but only the spectral energy distribution, which is then translated into stellar mass using models of stellar evolution. By setting the parameter of star formation efficiency $\varepsilon$ (the fraction of baryonic matter that has turned into stars), we can estimate the mass of baryons {in the observed galaxies}, and from it, through the cosmological ratio of baryonic and dark matter, the total mass. Such an estimate contains several sources of uncertainty: the efficiency of star formation can vary over a wide range, the fraction of baryons in a separate halo may differ from the average for the Universe. For this reason (since we do not have a reliable theory of galaxy formation that allows us to unambiguously find their parameters from visible spectra), it is still impossible to draw unambiguous conclusions about the discrepancy between observations and $\Lambda$CDM cosmology. Attempts to test cosmology using these observations are still limited to restrictions on $\varepsilon$, for example, \cite{Xiao23} found three massive galaxies at $z\sim5-6$, which in the $\Lambda$CDM model require an unrealistic value of $\varepsilon>0.2$.

In the works \cite{Padmanabhan_2023,Tkachev23} it was shown that adding a peak to the power spectrum can reduce this possible contradiction between observations and cosmology, reducing the value of $\epsilon$ needed to explain the observations. Thus, according to \cite{Tkachev23}, in the model with a peak for all galaxies considered in this article, $\epsilon<0.1$ is sufficient, which is consistent with known data on the efficiency of star formation  \citep{Giodini_2009,Silk2018,Lovell22}. {Qualitatively, we can say that adding a peak to the spectrum leads to the appearance of a new population of objects -- compact clumps that were formed earlier than halos in the $\Lambda$CDM model and are numerous at large $z$.}

{As Zeldovich showed \citep{Zeldovich70}, inhomogeneities in an expanding self-gravitating dusty medium evolve not only into more or less spherical clumps (halos), but also form a network of one-dimensional filaments and two-dimensional walls. For Gaussian perturbations, this conclusion qualitatively does not depend on the shape of the spectrum \citep{Doroshkevich70}. Due to the difficulties with measuring the halo mass function in observations, mentioned above, we propose to investigate the quantitative parameters of walls and filaments, which, perhaps, can be restored in observations from the spatial distribution of galaxies.}

{Previously obtained results from \cite{Tkachev23} showed that in models with a peak in the spectrum, already at $z\sim25$, the first halos are formed with a mass of $10^8$~$h^{-1}M_\odot$ (<<superdense clumps>>). At $z\sim10$, where current observations reach, the number of halos with masses of $10^9-10^{10}$~$h^{-1}M_\odot$ is several times greater than the number of halos in the $\Lambda$CDM model. At lower redshifts, due to the nonlinear evolution of the large-scale structure, this excess of halos practically disappears, and by $z\sim0$, the differences in the mass function are only tens of percent. Thus, the observational manifestations of peaks in the spectrum need to be sought precisely at large redshifts, where they are more clearly visible.}

Changing the power spectrum of perturbations not only changes the mass function, but also gives a number of other predictions that, in our opinion, will allow these models to be tested: changes in the large-scale structure and changes in the structure of halos (and galaxies) due to their earlier formation. These two predictions are analyzed in this work. To study the nonlinear evolution of the structure, we performed numerical calculations of the solution of the many-body problem for a dark matter medium (cosmological simulations). Due to the fact that the processes of galaxy formation at $z>10$ are not modeled reliably enough now, we consider only dark halos. Not all of them can contain galaxies, also the luminosity of a galaxy does not have to be proportional to the mass of the halo, which leads to some observational selection of halos. For simplicity, in our work, we consider all halos with a mass above a certain threshold, we leave the modeling of galaxy formation and selection processes for future research.

The spatial distribution of galaxies or halos can be described quantitatively using the minimal spanning tree (MST) method \citep{Barrow85,DTAW04,Demianski11}. This method connects objects (halos or galaxies) in space with segments so that one can <<travel>> from one object to any other in a unique way, and the total length of the segments is minimal. The distribution of the lengths of the MST segments is one of the characteristics of the spatial distribution, in particular, it allows to distinguish systems of objects, distributed randomly in three-dimensional space, from those strung on randomly located {two-dimensional surfaces}, or {one-dimensional lines} (which resembles a network of pancakes or filaments of large-scale structure). By setting a certain threshold length of the segment and discarding all longer segments, we can obtain a system of clusters -- regions of increased density of objects. These clusters can also be divided into filaments and pancakes according to some signs \citep{Doroshkevich01,demianski04}. In our work, we also apply the MST method to study the clustering of halos at $z\sim10$.

For describing the internal structure of halos, we adhere to the concept outlined in the article \cite{Demianski23}. The halo is approximately considered as a spherically symmetric system in which matter is distributed according to the Navarro-Frenk-White (NFW) density profile. This profile is characterized by two parameters, which are convenient to use as the maximum circular velocity $v_{max}$ and the value defined in \cite{Demianski23} $w \equiv v_{max}/r_{max}$, where $r_{max}$ is the radius at which the maximum circular velocity is reached. For comparison, we also use another description popular in the literature in terms of virial mass $M_v$ and concentration parameter $c \equiv r_v/r_{s}$, where $r_v$ is the virial radius of the halo (the radius within which the average density is $\Delta \approx 200$ times higher than the critical density of the Universe, which corresponds to the satisfaction of the virial relation for the spherical model \cite{Bryan98}), and $r_s\approx r_{max}/2.2$ is the characteristic radius of the NFW density profile inflection.

\section{NUMERICAL MODELS}
Various scenarios of cosmological inflation lead to the formation of spectra of complex shape, sometimes including numerous peaks on different scales, \citep[see, for example, ][]{Ivanov94}. To avoid being tied to a specific model of inflation, we describe the peak with the simplest form -- a Gaussian curve (in logarithmic coordinates), which is characterized by the position of the center of the peak $k_0$, its amplitude $A$ and width $\sigma_k$:

\begin{eqnarray}
    & P_\mathrm{bump}(k) = P_\mathrm{\Lambda CDM}(k) \times \nonumber \\
    & \times \left( 1 + A \exp \left( -\frac{(\log(k)-\log(k_0))^2}{\sigma_k^2} \right) \right).
    \label{eq:bump}
\end{eqnarray}

\begin{table*}
\caption {Parameters of numerical models} \label{tab:sims}
\medskip
\begin{tabular}{c|c|c|c|c|c}
\hline
Model        & $\Lambda CDM$ & k7 & k15 & k30 & k80 \\
\hline
Cube size (Mpc/$h)$       & 5.0           & 5.0        & 5.0        & 5.0        & 5.0    \\
Number of test particles $N_{total}$ & $512^3$       & $512^3$    & $512^3$    & $512^3$    & $512^3$   \\
Initial redshift         & $10^3$         & $10^3$     & $10^3$     & $10^3$     & $10^3$    \\
Final redshift        & $8$           & $8$        & $8$        & $8$        & $8$       \\
Peak position $k_0$, $h/$Mpc                   & --            & 7          & 15         & 30         & 80        \\
Peak amplitude $A$                      & 0             & 20         & 20         & 20         & 10        \\
Peak width $\sigma_k$                 & --            & 0.1        & 0.1        & 0.1        & 0.1       \\ 
\hline
\end{tabular}\\
\end{table*}

At the same time, for the case of narrow peaks with $\sigma_k<1$, physical responses, for example, the mass function, will depend only on the product of amplitude and width, i.e., on the integral under the curve (\ref{eq:bump}). Therefore, we further fix the width of the Gaussian $\sigma_k=0.1$ and change only its position and amplitude. 

In the work \cite{Tkachev23} an analysis of different amplitudes and positions of Gaussians was conducted, and it was shown that the variant with $k_0=7$~$h/$Mpc, $A=20$, $\sigma_k=0.1$ allows explaining a possible excess of galaxies at high redshifts under reasonable assumptions about the efficiency of star formation, $\epsilon<0.1$. In this work, we analyze this variant and all other peak variants for which numerical models were computed in \cite{Tkachev23}.
The parameters of the models are presented in Table~\ref{tab:sims}. Numerical models cover the range of the parameter $k_0$ from 7 to 80~$h/$Mpc and are named according to the value of the parameter $k_0$.

The parameters of the numerical models used are described in detail in the work \cite{Tkachev23}, here we provide a brief description. {The size of the model box and the number of particles in it are chosen so that all considered peak variants fit between the main mode of perturbations $k_1 = 2\pi/L$ and the Nyquist wave number $k_{Ny}=\pi/(L/N^{1/3})$. In addition, the main mode of perturbations should have an amplitude of no more than 1 at $z=8$ (end of the simulation), which is also fulfilled for $L=5$~Mpc/$h$.} The initial conditions for the models are set at $z=1000$ in the form of a random realization of a Gaussian field of peculiar velocities. The positions of the particles are obtained from the velocities in accordance with the approximate Zeldovich theory. At the same time, the same realization of the sequence of random numbers is used for all models, which eliminates random differences between the models. {The initial redshift was chosen from the condition that the Zeldovich approximation still works well at this moment for all models, i.e., the displacements of the particles from their initial positions do not exceed half the distance between neighboring particles.} The calculation of further evolution was performed by the GADGET-2 code \citep{gadget} up to a redshift of $z=8$.
Dark matter halos were identified using the Amiga Halo Finder code \citep{AHF} with the standard density criterion.

\begin{figure}
    \centering
    \includegraphics[width=\linewidth]{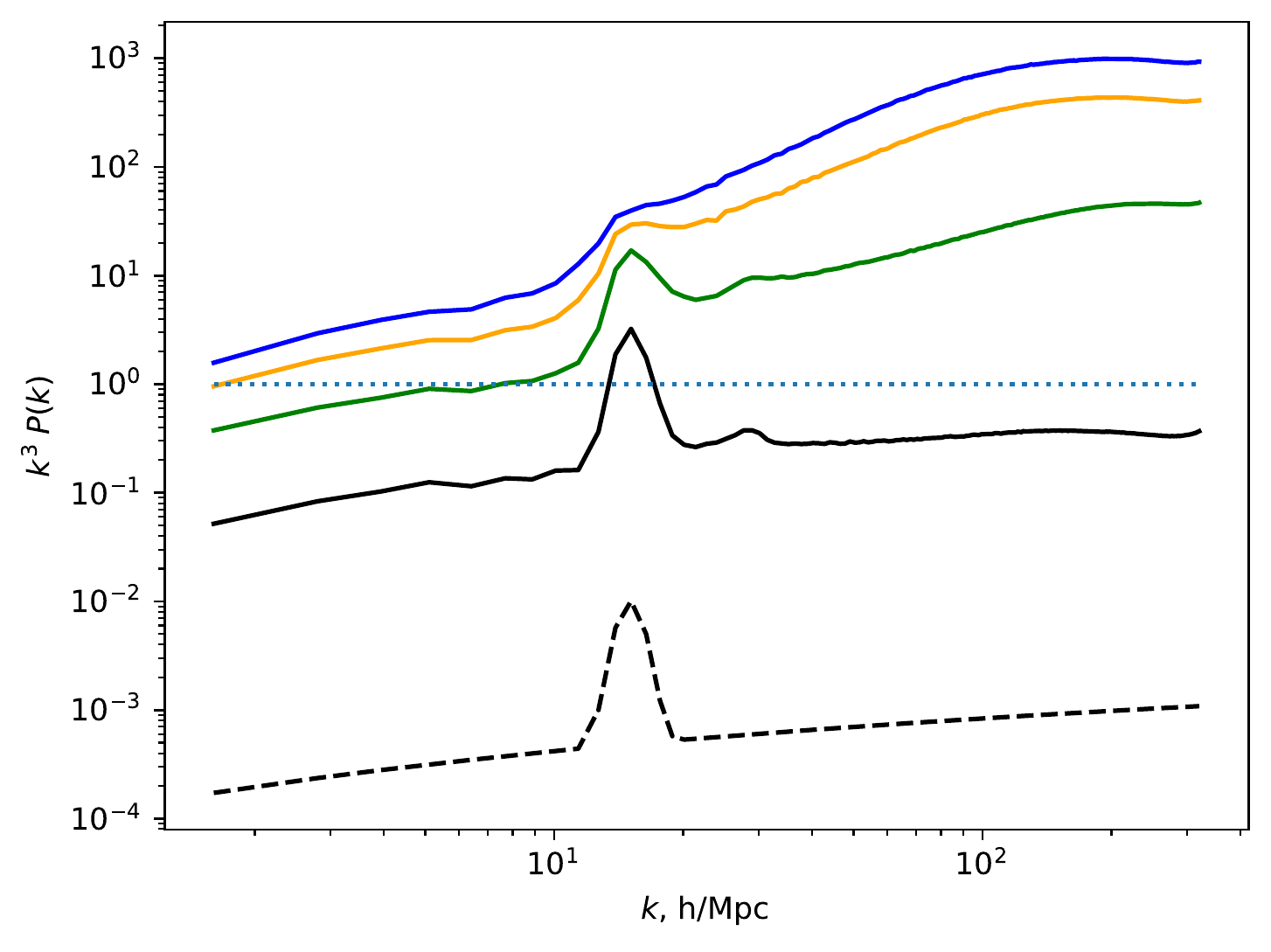}
    \caption{The power spectrum of density perturbations in the numerical model k15. Dashed line -- the initial spectrum at $z=1000$. Solid lines -- at $z=50$, 18, 11, 8 (bottom to top). Dotted line -- the boundary of linear and nonlinear perturbations.}
    \label{fig:Pk}
\end{figure}

An overall understanding of the evolution of matter distribution in the calculations can be obtained from Fig.~\ref{fig:Pk}. It shows the initial power spectrum of density perturbations and its change over time for the k15 model. The ordinate axis represents the dimensionless quantity $k^3 P(k)$ characterizing the amplitude of density perturbations on a scale of $2\pi/k$. Perturbations become nonlinear when this amplitude reaches unity. In this numerical model, this first occurs at $z\approx 100$ for perturbations with the scale of the peak, $k=15$~$h/$Mпк. At $z=36$ the first halos are formed with masses of $10^6-10^7$~$M_\odot/h$. At $z=18$, halos with a mass of $10^9$~$M_\odot/h$ appear, which roughly corresponds to the scale of the peak in the spectrum, and perturbations with $k>10$~$h$/Mпк become predominantly nonlinear. At $z=8$, perturbations on the scale of the cube become nonlinear, and there is no point in continuing further calculations.

\section{LARGE-SCALE STRUCTURE}
According to the theory proposed by Zeldovich \citep{Zeldovich70}, cosmological perturbations lead to the formation of a network of walls (pancakes), filaments and voids between them. An increase in power in a certain range of scales leads to an increase in density contrast and an earlier formation of this structure on the corresponding scales. In the k7 model, the peak on the spectrum is located on a linear scale of 0.9 Mpc/h, which at $z=10$ corresponds to an angular scale of 5.2 arc minutes, and in this model, one can expect the manifestation of observable structures from galaxies of such sizes. For comparison, the field of view of the most sensitive NIRCam camera on JWST is $2.25'\times 2.25'$, i.e., such scales are available in existing and planned surveys of distant galaxies. It is also interesting to note the recent work \cite{Wang23}, in which a filament about 10~Mpc in size was discovered at $z=6.6$, which suggests that the study of large-scale structure in the distribution of galaxies at large $z$ is a solvable problem.

\begin{figure*}
    \centering
    \includegraphics[width=0.9\textwidth]{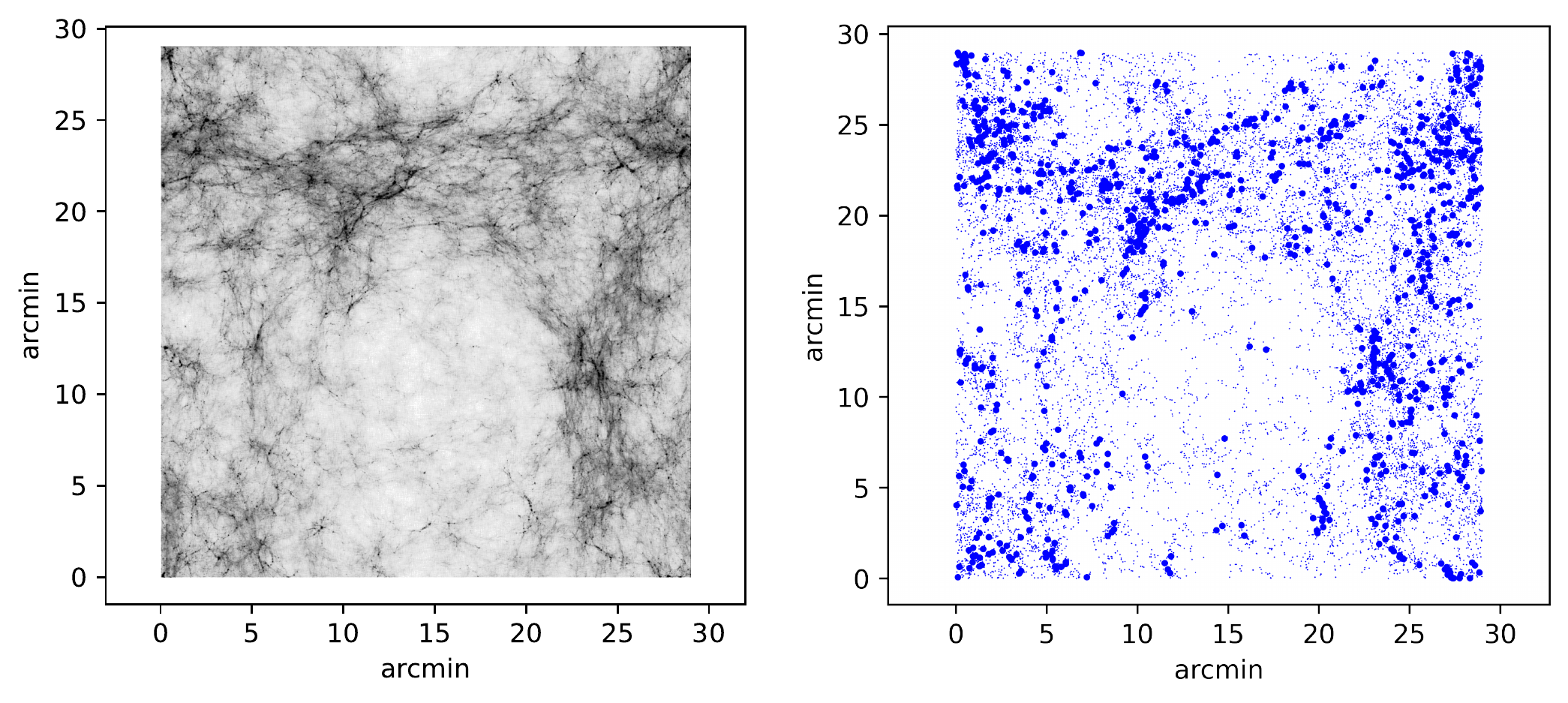}
    \includegraphics[width=0.9\textwidth]{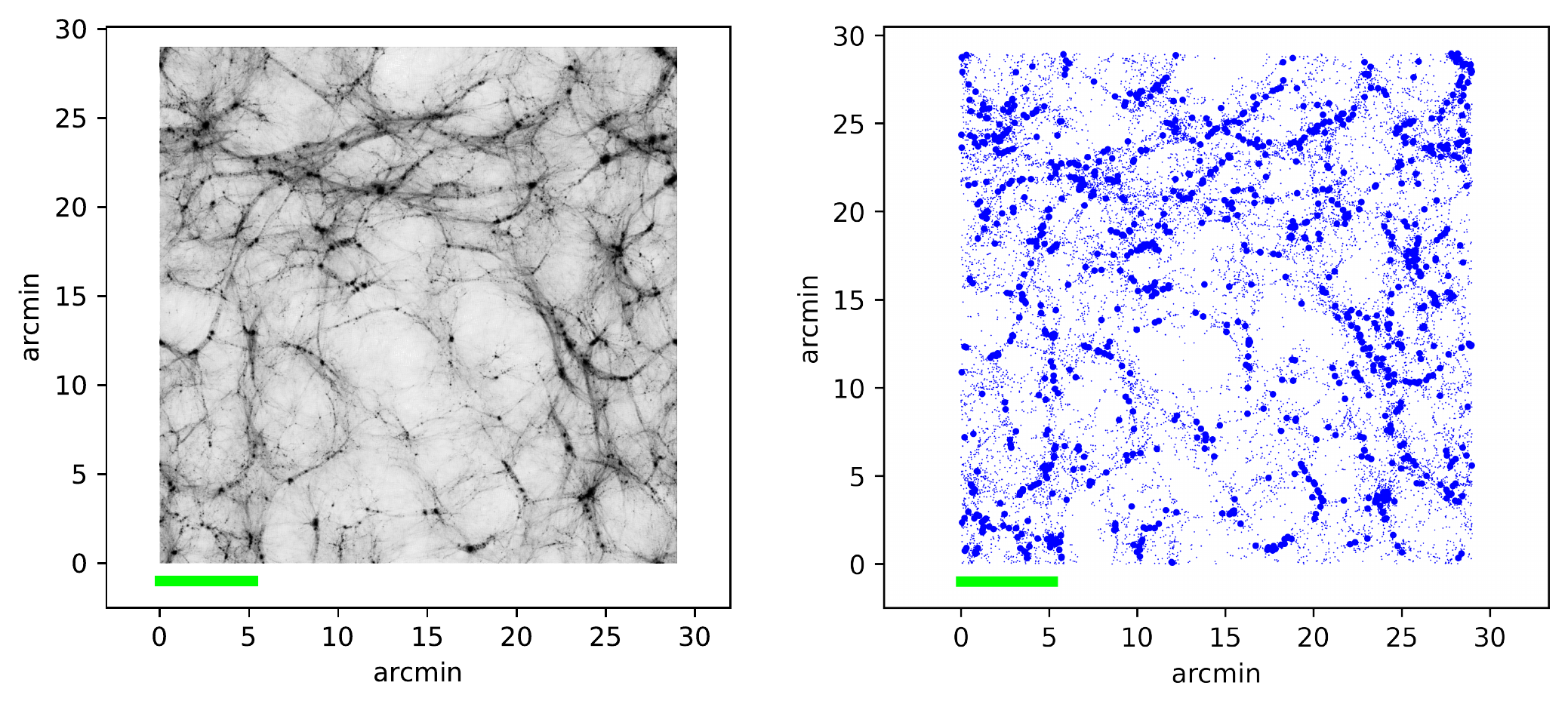}
    \includegraphics[width=0.9\textwidth]{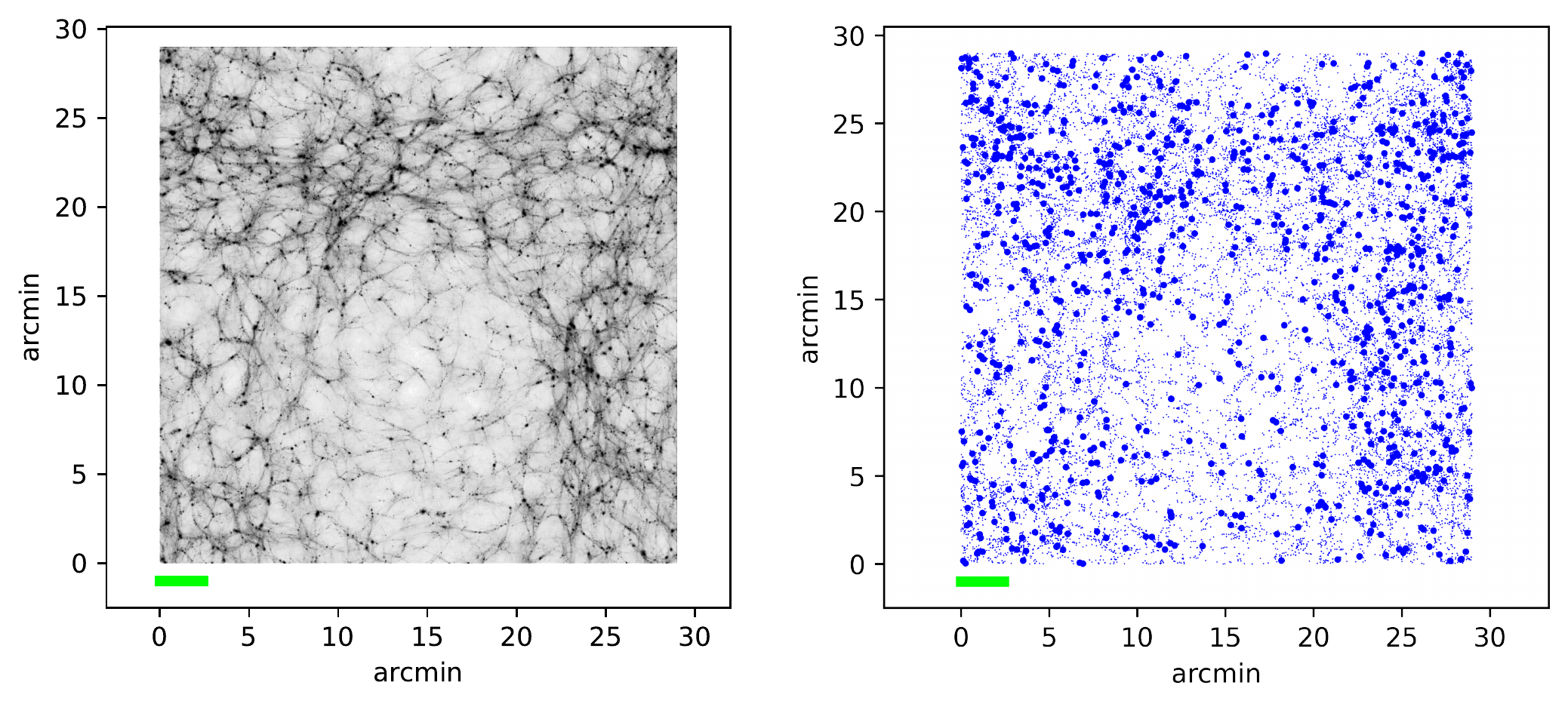}
    \caption{{Top to bottom: large-scale structure in the $\Lambda$CDM, k7, k15 models at $z=10$. Left: projection of the density of dark matter particles, right: dark matter halos. Large dots correspond to the 1000 most massive halos, small dots correspond to all the others (containing more than 20 particles). The horizontal line indicates the size corresponding to the peak scale $k_0$.}}
    \label{fig:lss}
\end{figure*}

\begin{figure*}
    \centering
    \includegraphics[width=0.9\textwidth]{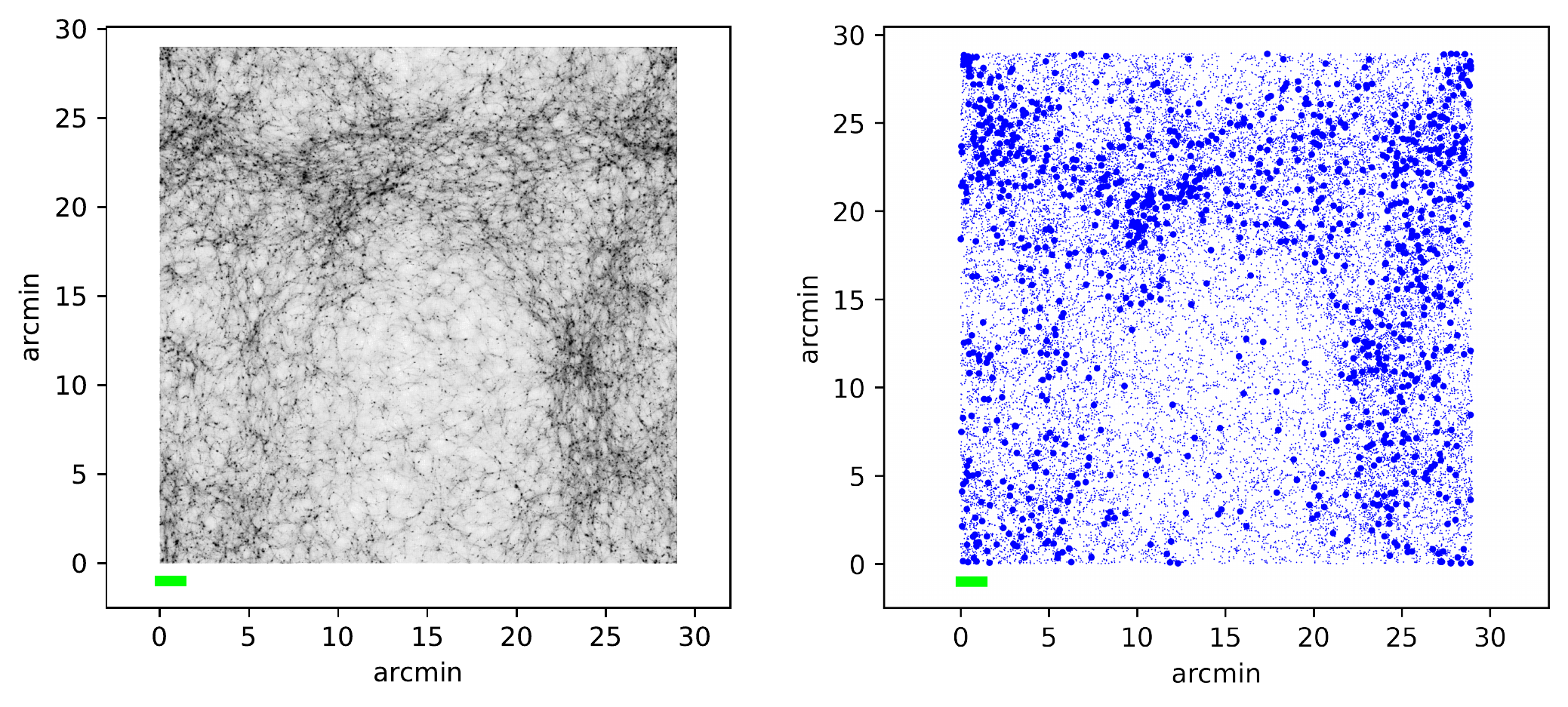}
    \includegraphics[width=0.9\textwidth]{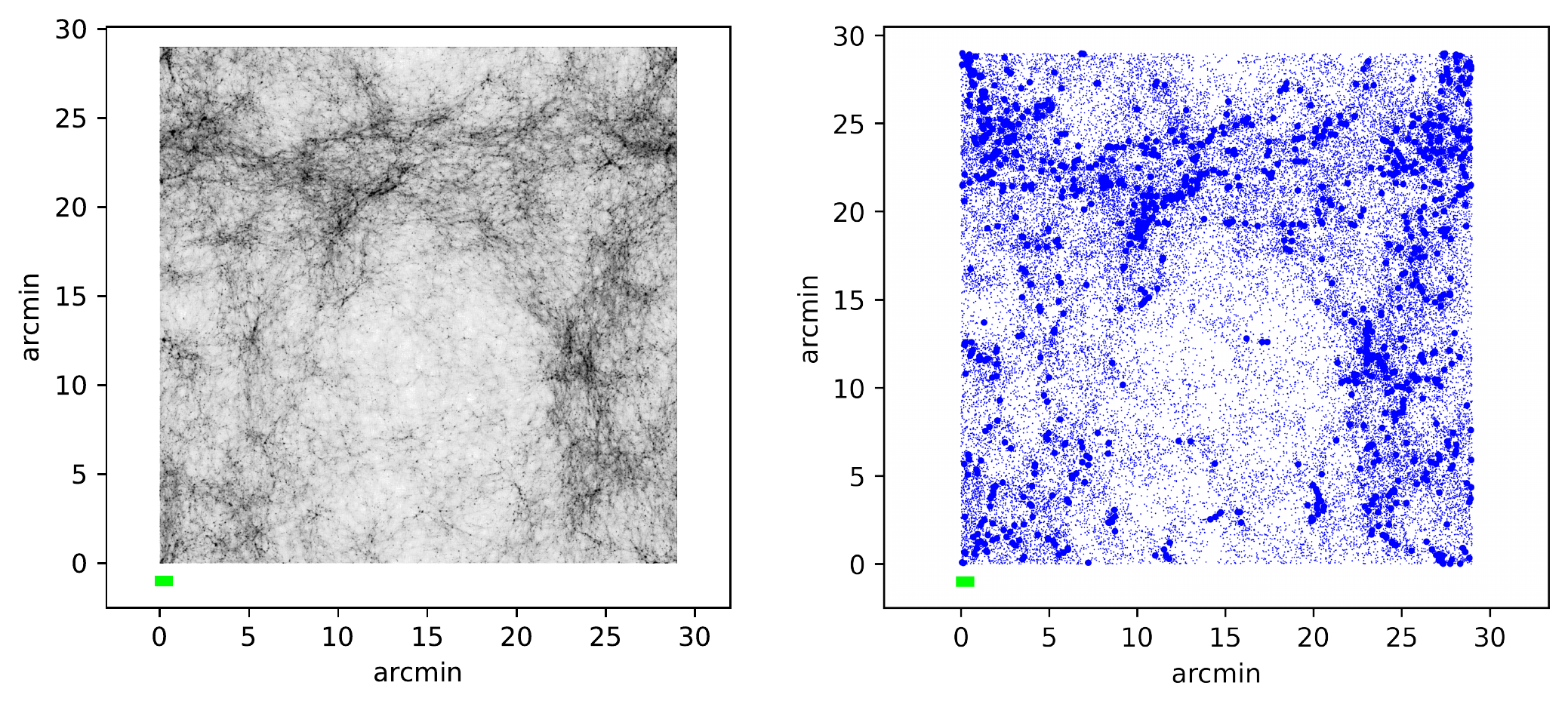}
    \caption{Continuation of Fig.~\ref{fig:lss} for the k30, k80 models.}
    \label{fig:lss2}
\end{figure*}

The type of large-scale distribution of galaxies for models with a peak in the spectrum is demonstrated in Fig.~\ref{fig:lss}-\ref{fig:lss2}. 1000 most massive halos from each model have been selected. The density of {massive halos in the right column} of Fig.~\ref{fig:lss}-\ref{fig:lss2} roughly corresponds to the observed density of the number of distant galaxies with $z\ge 10$ in existing JWST surveys: here 1.1 galaxies per square minute, 0.4 galaxies per square minute in the sample from the work \cite{Donnan23}, 1.1 galaxies per square minute in the survey from the article \cite{Bradley23}. However, in real surveys, a large range of redshifts is covered, while in Fig.~\ref{fig:lss} the depth of the image is only 5~Mpc/h. Nevertheless, this illustration clearly shows the difference between the $\Lambda CDM$ and k7 models. As $k_0$ increases, the differences gradually disappear: the cellular structure caused by the introduction of the peak shifts to the region of scales of individual galaxies.

{As can be seen from Fig.~\ref{fig:lss}-\ref{fig:lss2}, for different test objects, the introduction of peaks manifests itself differently. In the distribution of dark matter in all models with peaks, a characteristic ripple is visible on the scale of the peak, which is shown for clarity by a horizontal line. However, dark matter itself is not observable, so it is useful to consider how this ripple manifests itself in the distribution of halos of different masses (some halos may contain luminous objects). In the k7 model, the mass scale of the peak $M_0=(2\pi / k_0)^3 \rho_m=6\times10^{10}$~M$_\odot/h$ is significantly larger than the average mass of the first 1000 halos $10^{9}$~M$_\odot/h$. Therefore, in the k7 model, massive halos can be considered as test particles that clearly show the ripple due to the peak. In the k15 and k30 models, $M_0=6\times10^{9}$ and $8\times10^{8}$~M$_\odot/h$, respectively, which is already comparable to the average mass of the halos shown in the figure by bold dots, so the distribution of these objects looks blurred. At the same time, the cells with the scale of the peak can be traced in the distribution of halos of smaller masses. Finally, in the k80 model, the masses of almost all halos from the first thousand exceed the mass scale of the peak, so such halos no longer <<feel>> the presence of the peak, and their arrangement in space is practically identical to $\Lambda$CDM. But the peak in the k80 model leads to the formation of a large number of mini-halos, shown by small dots, the number of which is noticeably higher than in $\Lambda$CDM. Given that the manifestations of the peak depend on the sample of objects that we are looking at (on their masses), two samples are analyzed in this work: the first 1000 and the first 3000 most massive halos.}

\begin{figure}
    \centering
    \includegraphics[width=\linewidth]{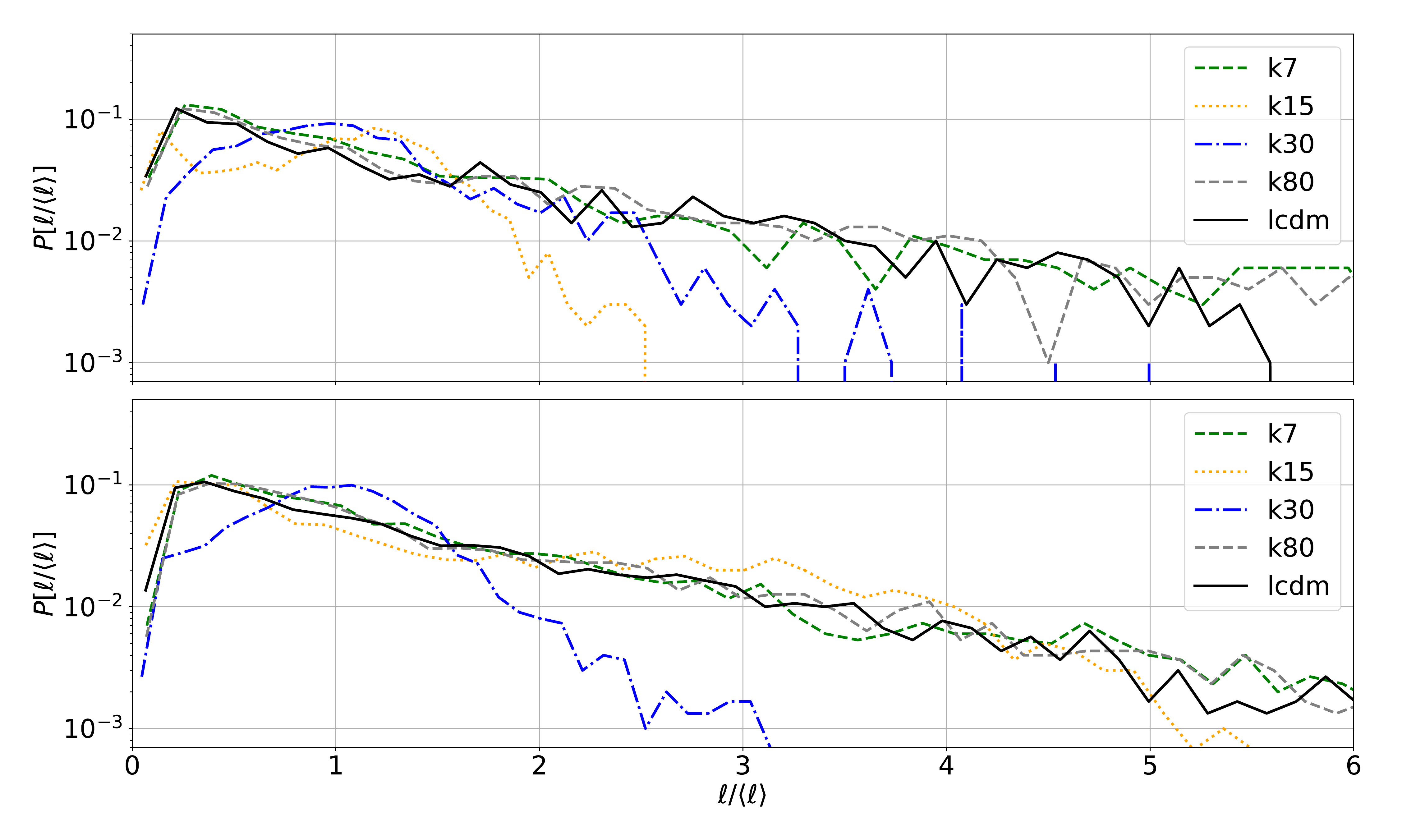}
    \caption{The distribution of the lengths of the MST segments for samples of 1000 most massive halos (top), and 3000 most massive halos (bottom panel) at $z=10$ for five cosmological models. {Table~\ref{tab:dist} shows the average parameters of the distributions}}
    \label{fig:ldist}
\end{figure}

To quantitatively describe the large-scale structure, we used the MST method, which gives several characteristics of the spatial distribution of objects. The first such characteristic is the distribution function of the lengths of the segments that make up the tree, $P(\ell)$. The average and median values of the length and their distribution functions for the 1000 and 3000 most massive halos from the considered models are presented in Table \ref{tab:dist} and in Fig.~\ref{fig:ldist}. {For most models, except for k15 and k30, the distribution functions at $\ell$ greater than some value have an exponential form, which corresponds to the theoretical expectation for the large-scale structure (such a distribution appears with the random arrangement of objects on one-dimensional lines and two-dimensional planes). The difference between the k15 and k30 models, apparently, is connected with the <<blurring>> of the large-scale structure mentioned above due to the coincidence of the Gaussian peak scale with the mass scale of the studied halos.}.

\begin{table}[]
    \centering
    \begin{tabular}{c|c|c|c|c}
         & $\langle \ell \rangle$ & $\ell_{med}$ & $\langle \ell \rangle$ & $\ell_{med}$ \\
         & kpc/$h$, & kpc/$h$, & kpc/$h$, & kpc/$h$, \\
         & $N=1000$ & $N=1000$ & $N=3000$ & $N=3000$ \\ \hline
        $\Lambda$CDM    & $174\pm5$ & $117$     &  $114\pm2$ & $79$ \\
        k7              &  $169\pm6$ & $94$    &  $92\pm2$ & $58$ \\
        k15              &  $288\pm5$ & $302$    &  $160\pm2$  & $109$ \\
        k30              &  $230\pm4$ & $200$    &  $182\pm2$ & $177$ \\
        k80              &  $163\pm5$ & $99$    & $109\pm2$ & $71$ \\ \hline
    \end{tabular}
    \caption{Average and median values of the lengths of the MST segments for samples of 1000 and 3000 most massive halos at $z=10$. {After $\pm$ the standard errors of the mean are given. Fig.~\ref{ldist} shows the distribution functions of the lengths of the segments.}}
    \label{tab:dist}
\end{table}

Another characteristic that can be obtained using the tree is the branching of the tree, which allows us to establish the proportion of filaments in the structure. To study this value, the tree is first broken down into clusters by discarding all segments with a length greater than some $\ell_{thr}$, and also removing all clusters containing less than $M_{thr}$ objects. Then each cluster is divided into a <<trunk>> -- a set of segments that form the longest one-way path along the tree, and <<branches>> -- segments that did not fall into the trunk. The ratio of the trunk length $L_{trunk}$ to the total length of the segments of this cluster (trunk + branches) $L_{tree}$ characterizes the degree of elongation of the cluster \citep{Doroshkevich01}. Thus, filamentous structures should have $L_{trunk}/L_{tree}\approx 1$. 

We used $\ell_{thr}=  \ell_{med} $, $M_{thr}=5$. This choice is due to the fact that with a minimum mass of 3 or less, it is impossible to split part of the tree into a trunk and branches, and the larger the mass of the cluster, the more original objects will have to be excluded from the analysis. The threshold length affects the number of clusters: if it is small, a significant part of the objects is thrown out of the analysis, if it is too large, the objects are combined into one or more large clusters (percolation occurs) and it is pointless to talk about the shape of such clusters. Small changes in the threshold parameters (for example, $M_{thr}=6$, $\ell_{thr}=1.2 \ell_{med}$) do not qualitatively affect the conclusions we have reached. 

Consider a typical cluster with $M=5$ points for our criterion. If they all form a trunk, obviously $L_{trunk}/L_{tree}=1$, but if there is one branch, then $L_{trunk}/L_{tree}=0.8$. Therefore, we will call filaments clusters with $L_{trunk}/L_{tree}\geq 0.8$.

\begin{figure}
    \centering
    \includegraphics[width=\linewidth]{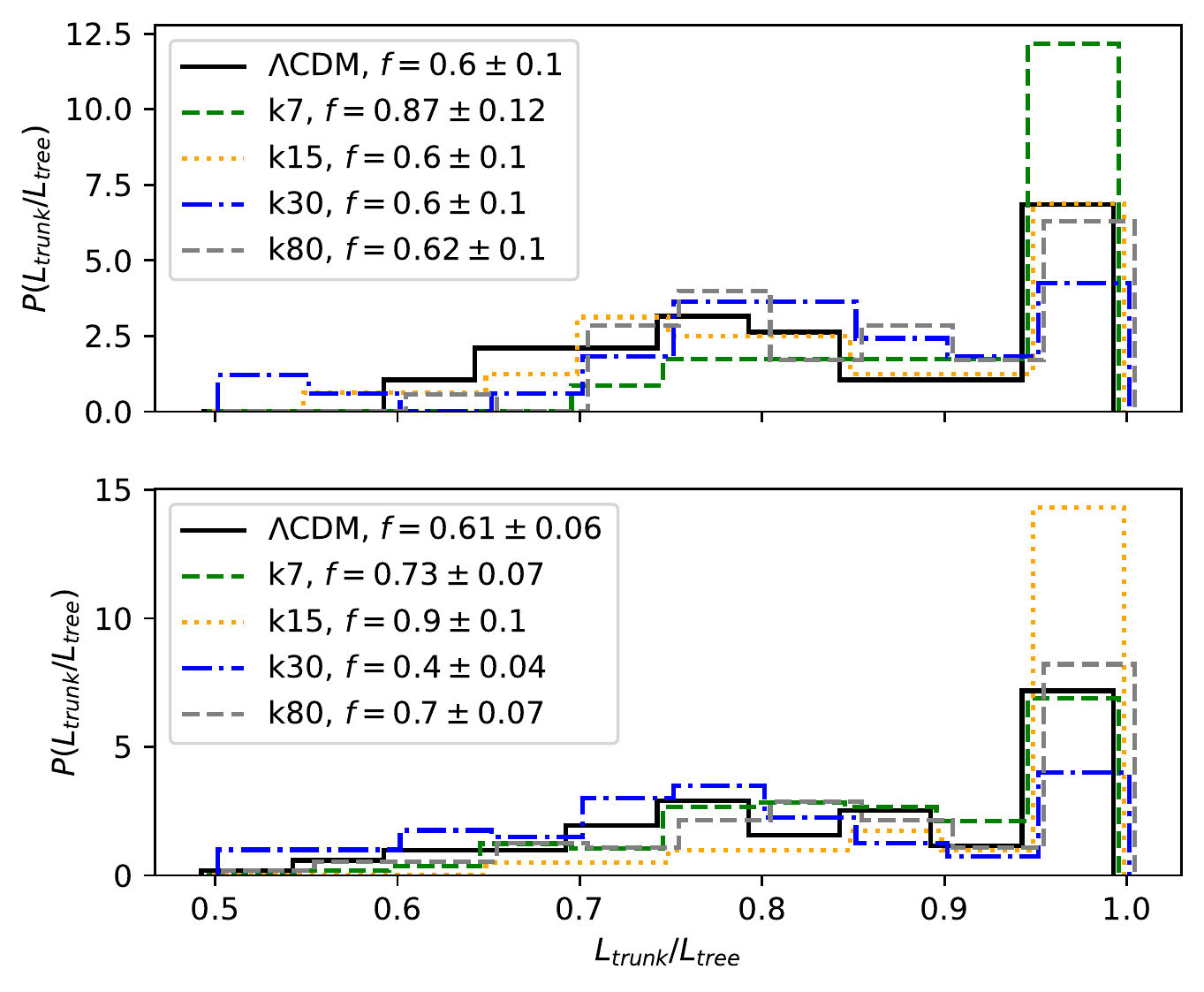}
    \caption{The distribution of the ratio of the length of the <<trunk>> of the tree to the total length of the tree $L_{trunk}/L_{tree}$ for five cosmological models for samples of 1000 halos (upper graph) and 3000 halos (lower graph) {at $z=10$}. The histograms are slightly shifted along the horizontal axis to avoid overlapping lines.}
    \label{fig:trunk-tree}
\end{figure}

The distributions of the ratio $L_{trunk}/L_{tree}$ are presented in Fig.~\ref{fig:trunk-tree}, the legend indicates the proportion of filaments $f$ (clusters with $L_{trunk}/L_{tree}\geq 0.8$). Based on these data, we can draw the following conclusions: in the k7 model for a sample of 1000 halos and in the k15 model for a sample of 3000 halos, there is an increased number of filaments with $L_{trunk}/L_{tree}>0.8$ compared to the $\Lambda$CDM model, in the k30 model for a sample of 3000 halos -- reduced, and in the k80 model comparable to $\Lambda$CDM. The dependence of the result on the sample of halos demonstrates the fact that by the distribution of halos one can see only the filaments of the large-scale structure, which are represented by a sufficiently large number of halos.

\section{STRUCTURE OF HALO}
The work \cite{Demianski23} presents a model for describing dark matter halos and compares halos from numerical models with observed galaxies. This model characterizes the halo by two parameters: the maximum circular velocity $v_{max}$ and the density parameter $w$\footnote{In the article \cite{Demianski23}, the value of $w$ was denoted by the letter $h$. Characteristic density inside $r_{max}$: $1/4\pi Gw^2$.}:

\begin{eqnarray}
    v_{max}^2 &=& max\left( \frac{GM(r)}{r} \right) \\
    w &=& \frac{v_{max}}{r_{max}}.
\end{eqnarray}

This description is slightly different from that adopted in the literature on cosmological simulations, where two other characteristics are introduced: the virial mass $M_v$ and the concentration parameter $c$ \citep[see, for example][]{Klypin_2011}. The description in terms of $c$, $M_v$ is convenient for simulations, but in observations, the virial radius is difficult to measure, since it is located on the periphery of the halo, where there is practically no visible matter. In addition, these parameters correlate with each other, so any selection effects that affect one parameter also affect the distribution of the second parameter. For example, we cannot talk about the distribution of the concentration parameter if we do not specify what range of virial mass we are talking about.

From the point of view of theory or simulations, these two descriptions are equivalent, since by setting the NFW density profile, we can switch from the parameters $c$, $M_v$ to the parameters $w$, $v_{max}$ or vice versa. In the work \cite{Demianski23}, the description of $w$, $v_{max}$ was applied to observed galaxies, and it was shown that the parameters measured in observations for typical galaxies at $z\sim0$ correspond to the parameters of halos from simulations, which demonstrates the applicability of such a description for comparing observations and theory.

\begin{figure*}
    \centering
    \includegraphics[width=0.9\textwidth]{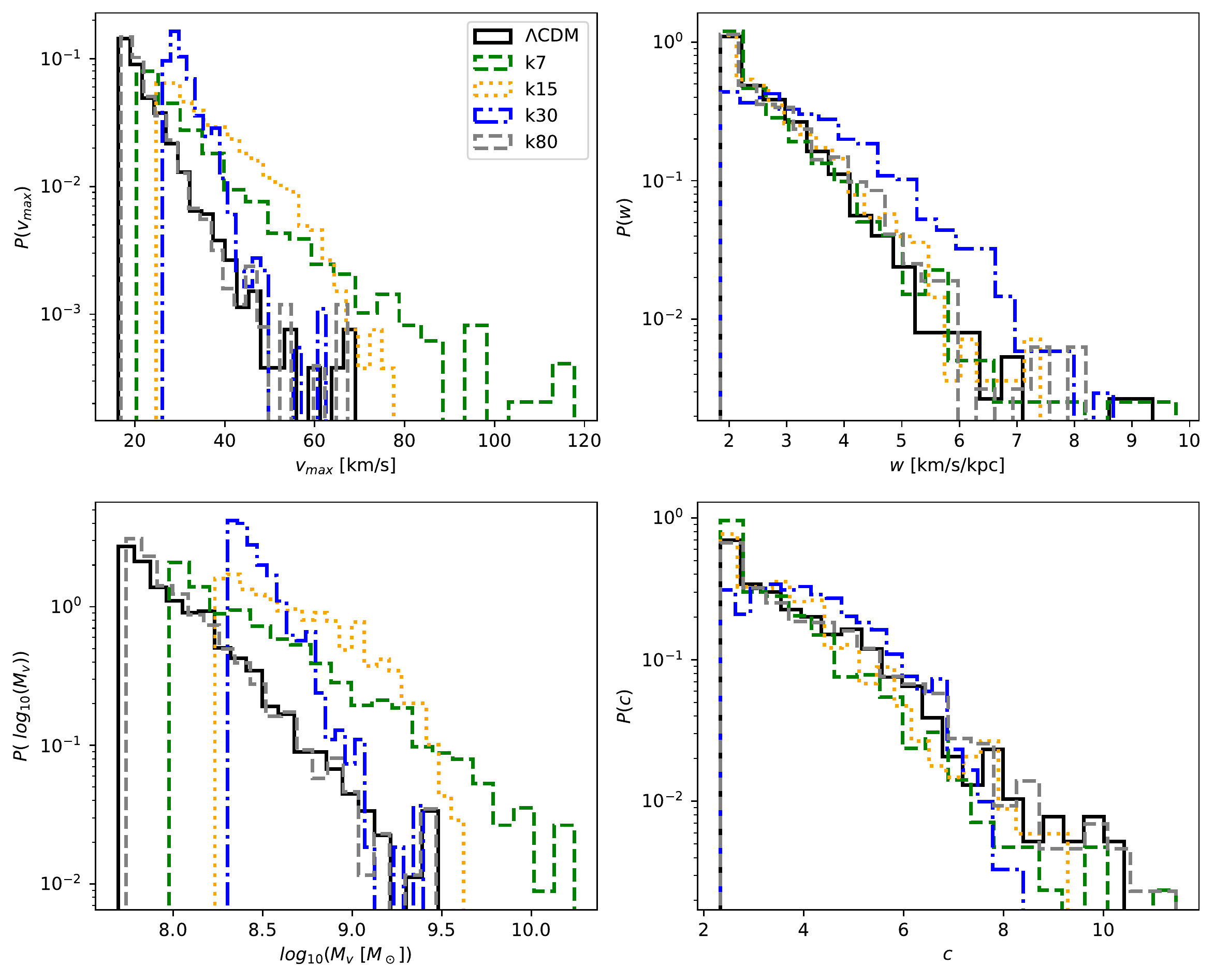}
    \caption{Histograms of the distribution of parameters of the 1000 most massive halos  {at $z=10.1$} in 5 numerical models. Solid black line -- model $\Lambda CDM$, black dashed -- k7, black dotted -- k15, black dash-dotted -- k30, gray dashed -- k80. {The parameters of the distributions are presented in Table~\ref{tab:params}.}
    \textit{Top left panel:} maximum circular velocity $v_{max}$. 
    \textit{Top right panel:} halo density parameter $w = v_{max}/r_{max}$.
    \textit{Bottom left panel:} logarithm of virial mass $M_v$.
    \textit{Bottom right panel:} concentration parameter $c = r_v/r_{s}$.}
    \label{fig:conc}
\end{figure*}

\begin{table*}[]
    \centering
    \begin{tabular}{c|c|c|c|c|c}
    \hline
       Model  & $\Lambda CDM$ & k7 & k15 & k30 & k80  \\ \hline
       Median $v_{max}$, km/s & 20.1 & 27.6 & 33.7 & 30.0 & 20.5 \\
       Average $v_{max}$, km/s  & $22.2\pm0.2$ & $31.9\pm0.4$ & $36.3\pm0.3$ & $31.2\pm0.1$ & $22.4\pm0.2$ \\
       Dispersion $v_{max}$, km/s & 6.3 & 13.4 & 9.6 & 4.2 & 6.1 \\ \hline
       Median $w$, $h$km/s/kpc &  2.4 & 2.3 & 2.5 & 3.1 & 2.5 \\
       Average $w$, $h$km/s/kpc & $2.6\pm0.03$ & $2.6\pm0.03$ & $2.7\pm0.03$ & $3.3\pm0.03$ & $2.7\pm0.03$ \\
       Dispersion $w$, $h$km/s/kpc & 0.9 & 1.0 & 0.9 & 1.1 & 0.9 \\ \hline
       Median $M_v$, $10^8\;M_\odot/h$ & 0.8 & 2.1 & 3.9 & 2.7 & 0.9 \\
       Average $M_v$, $10^8\;M_\odot/h$ & $1.4\pm0.07$ & $5.7\pm0.4$ & $6.0\pm0.2$ & $3.2\pm0.06$ & $1.4\pm0.07$ \\
       Dispersion $M_v$, $10^8\;M_\odot/h$ & 2.2 & 13.4 & 5.5 & 1.8 & 2.1 \\ \hline
       Median $c$ & 3.4 & 3.0 & 3.4 & 4.0 & 3.4 \\
       Average $c$ & $3.8\pm0.04$ & $3.4\pm0.04$ & $3.6\pm0.04$ & $4.1\pm0.04$ & $3.9\pm0.05$ \\
       Dispersion $c$ & 1.4 & 1.2 & 1.3 & 1.2 & 1.5 \\ \hline
       
    \end{tabular}
    \caption{Parameters of distributions $v_{max}$, $w$, $M_v$, and $c$ at $z=10.1$ in the numerical models studied. {The distributions are shown in Fig.~\ref{fig:conc}}}
    \label{tab:params}
\end{table*}

The value of $v_{max}$ characterizes the depth of the potential well created by the dark halo, which, in turn, is important for the first galaxies: the gas accreting onto the halo is heated to a temperature of the order of
\begin{equation}
    T \approx \frac{\mu m_p v_{max}^2}{2k_b},
\end{equation}
where $m_p$ is the mass of the proton, $k_b$ is the Boltzmann constant, $\mu$ is the average molecular weight of the gas, $\mu=1.22$ for neutral primary gas. If the halo has $v_{max}>12$~km/s, then the gas in it will heat up to $10^4$~K and ionize, which will allow the gas to effectively relieve energy and, as a result, form stars \citep[see, for example, ][]{BarkanaLoab01,Haiman2000a}. The larger $v_{max}$, the more difficult it is to throw gas out of the galaxy as a result of supernova explosion processes, therefore, in halos with a large $v_{max}$ there is more gas left for further star formation, and such halos can potentially contain galaxies with a larger stellar mass.

The value of $w$ characterizes the density of matter inside the halo, the higher this value, the greater the density. The mean density inside $r_{max}$ is $1/4\pi Gw^2$. Obviously, with the same $v_{max}$ for two halos, $w$ will be larger for the one for which the maximum circular velocity is reached at a smaller radius, i.e., halos with large $w$ are more compact.

We selected 1000 of the most massive halos from each of our numerical models. Histograms of the distribution of parameters $v_{max}$, $w$, $M_v$, $c$ are presented in Fig.\ref{fig:conc}, and their medians, average values, and dispersions are given in Table~\ref{tab:params}. From these distributions, several conclusions can be drawn:

\begin{enumerate}
    \item In all models, the considered halos have $v_{max}>12$~km/s, which means that galaxies can arise in them.
    \item Models k7, k15, and k30 demonstrate an excess of $v_{max}$ velocities (as well as halo mass $M_v$) compared to $\Lambda$CDM. Galaxies in these models should also possess larger stellar masses than in $\Lambda$CDM. If it becomes possible to estimate the circular velocities of distant galaxies using spectroscopy, this can be used to test models with a peak at $k_0\le 15$~$h$/Mpc: in such models, galaxies with $v_{max}>100$~km/s may already be encountered at $z=10$, while in $\Lambda$CDM, the detection of such galaxies is unlikely.
    \item The distribution of the parameter $w$ shows that only in the k30 model there is a significant increase in density inside the halo, meaning they are more compact in this model than in other models. This is due to the fact that the scale of the peak corresponds to the typical masses of these halos: from $2\times 10^8$ to $2\times10^9$~$M_\odot/h$. 
    \item In the k80 model, the halo parameters from the considered sample do not differ from those in $\Lambda$CDM.
\end{enumerate}

\section{CONCLUSION}

This work demonstrates the possibility of studying the small-scale power spectrum of initial density perturbations by measuring the internal structure and spatial distribution of galaxies at high redshifts. Deviations of the spectrum from the standard one (from the $\Lambda$CDM model) were set in the form of a Gaussian with a variable position. {In the proposed model, the peak in the spectrum allows dense clumps to form earlier than in the $\Lambda$CDM model, which qualitatively explains the observations of galaxies at $z>10$ by the JWST telescope \citep{Naidu2022b,Castellano22,Finkelstein22,Donnan23,Labbe23}, as well as, perhaps, the population of <<little red dots>> (LRD) \citep{Matthee24,Akins24}. These are compact and numerous objects at redshifts $z=3-9$, whose nature is still unknown.}

We have performed numerical calculations of the formation of large-scale structure and dark matter halos for five models: $\Lambda$CDM and 4 models with peak positions in the power spectrum at $k_0=7$, 15, 30, 80~$h/$Mpc. {The amplitudes of the peaks were selected based on the fact that the number of dense clumps (halos) arising in the model with a peak at large redshifts should significantly exceed the number of halos of the same masses in $\Lambda$CDM.}

Samples of halos from these models demonstrate the dependence of the large-scale object distribution on the position of the peak: for models with $k_0=7$ and $k_0=15$~$h/$Mpc, using the minimal spanning tree method, an increased number of filaments is detected. Also, in the model with $k_0=30$~$h/$Mpc, the distribution of branch lengths of the tree noticeably differs from the $\Lambda$CDM model, and the number of filaments in the distribution of massive halos is significantly lower.

Models with $k_0=7$, 15, 30~$h/$Mpc demonstrate higher values of the maximum circular velocity of halos ($v_{max}$) compared to $\Lambda$CDM, as well as somewhat more compact halos (with larger values of the parameter $w$) in the model with $k_0=30$~$h/$Mpc. The model with $k_0=80$~$h/$Mpc, based on the parameters we analyzed, does not differ from the $\Lambda$CDM model.

In the future, it is necessary to move from the analysis of dark halos to models of galaxies, for which one can use both simple semi-analytical models and hydrodynamic simulations with star formation. Having catalogs of model galaxies, it will become possible to build cones of model surveys for JWST, Millimetron, SKA, ALMA observatories, taking into account the peculiarities of galaxy selection by a particular observatory. As a result, it will be possible to provide predictions for observations on the expected parameters of the large-scale structure and galaxy parameters for models with the $\Lambda$CDM perturbation spectrum and models with a modified spectrum.

Also, although we set deviations from the $\Lambda$CDM model in the form of distortions of the power spectrum in the range $7<k<80$~$h/$Mpc, which corresponds to scales $\lambda = 2\pi/k = 0.08-0.9$~Mpc/$h$, this is not the only possible way to set additional small-scale perturbations in the same range of scales. If we abandon the requirement of Gaussianity of initial perturbations at small scales, they can be set in the form of separate density peaks (perturbations in r-space, not in k-space). It can be expected that with peak sizes in the range $0.1-1$~Mpc/$h$, they will be able to form galaxies and supermassive black holes earlier than in the $\Lambda$CDM model (depending on the amplitude of the peaks). Similar models can also be investigated for their testability by observations using simulations.

\section{Acknowledgments}
The authors are grateful to T.I. Larchenkova for moral support and D.I. Novikov for valuable discussions.

\section*{FUNDING}
The work was supported by the LPI project NNG 41-2020.

\section*{CONFLICT OF INTEREST}
The authors declare no conflict of interest.

\bibliographystyle{aspb1}
\bibliography{bump}
\onecolumngrid
\clearpage

\end{document}